\documentstyle[prl,aps,epsf,epsfig,multicol]{revtex}

\newcommand{\be}{\begin{equation}}
\newcommand{\ee}{\end{equation}}
\newcommand{\ba}{\begin{array}}
\newcommand{\ea}{\end{array}}

\begin{document}

\title{Transition to Localization of Biased Walkers in a Randomly Absorbing 
Environment}

\author{Vishal Mehra and Peter Grassberger}

\address{John-von-Neumann Institute for Computing, Forschungszentrum J\"ulich, 
D-52425 J\"ulich, Germany}

\date{\today}

\maketitle

\begin{abstract}
We study biased random walkers on lattices with randomly dispersed 
static traps in one, two and three dimensions. As the external bias is increased 
from zero the system undergoes a phase transition, most clearly manifested in the 
asymptotic drift velocity of survivors which is zero below a critical bias. This 
transition is second-order in one dimension but of first order in higher 
dimensions. The model can be mapped to a stretched polymer with attractive 
interaction between monomers, and this phase transition would then describe 
sudden unfolding of the polymer when the stretching force exceeds a critical 
value. We also present precise simulations of the zero bias case where we 
show unambiguously that the transition between the Rosenstock and 
Donsker-Varadhan regimes is first order in dimension $\ge 2$.
\end{abstract}

\begin{multicols}{2}

\section{Introduction}

The behavior of a particle diffusing amidst randomly dispersed static traps is already
interesting by itself, but also because of its connections to several other problems:
It is closely related to the Anderson model of delocalization and to some models 
of population dynamics \cite{nelson}. It can be mapped exactly onto flux lines 
in a medium with parallel columnar defects \cite{nelson} and to the collapse of 
a self-attracting polymer without excluded volume interaction.

The problem is usually formulated as a random walk on a $d-$dimensional lattice; a 
random fraction $c$ of all lattice sites is occupied by traps which kill a walker 
if it happens to land on them. The main quantity of interest is the survival 
probability after $t$ steps, $P(c,t)$, averaged over all walks and all trap 
configurations. For small $t$ and small trap concentrations, $P(c,t)$ decays
exponentially as predicted by Rosenstock (RS) \cite{r70}, but at later times it 
crosses over to the exact Donsker-Varadhan (DV) form \cite{dv75,gp82,kh83}
\be
   \ln P(c,t) \sim -A_d \lambda^{\frac{2}{d+2}}t^{\frac{d}{d+2}} \label{DV}
\ee
where $\lambda=-\ln(1-c)$ and $A_d$ are exactly known numerical constants. Eq.(\ref{DV})
can be understood in terms of contributions from large trap-free regions. Such 
regions are very rare, but survival is much enhanced in them. Thus, $P(c,t)$ is 
dominated by regions whose size grows with $t$ as $R \sim t^{1\over d+2}$ 
\cite{gp82}.

The precise form of this crossover has been a matter of some controversy 
\cite{kzb84,f84,bhkgs97,n89}. In particular, simulations have been notoriously difficult 
for $d\ge 2$. While deviations from the RS prediction are easily seen \cite{gp82}, it has 
proven difficult to access the validity of  Eq.(\ref{DV}) in  simulations
\cite{kzb84,f84,bhkgs97}. For $d=3$, the situation has largely been clarified by
simulations of Anlauf \cite{anlauf} which unfortunately  were never published, and by
analytical work by Nieuwenhuizen \cite{n89}. Very recently, similar scaling was
found in \cite{gak01,bbb01} to hold also for $d=2$.

In this paper we first present further simulations of this model where we correct
some previously made claims and verify the prediction of \cite{n89} that the 
RS-DV crossover is akin to a first order phase transition at infinite temperature.
We then show results of the model with a superimposed bias. In the latter, the jumping 
rates in the (fixed) bias direction are enhanced by a factor $b>1$ and those in the 
opposite direction are suppressed by the same factor.

This external bias leads to a phase transition in the asymptotic dynamics. For small 
bias the long time dynamics is similar to the 
zero-bias case: the walkers stay in  large trap-free regions and do not drift.
Instead, their typical positions increase slowly as $x\sim R \sim t^{1/(d+2)}$. But 
above a critical bias $b_c$, the surviving walks drift with a finite velocity \cite{footnote}.
These behaviors are separated by a sharp transition which is of 
second-order in $d=1$ but of first order in $d\ge 2$. At $b\searrow b_c$ the 
drift velocity tends to zero in $d=1$ but reaches a non-zero value $v_c$ in 
$d\ge 2$. Asymptotically the survival probability decays 
exponentially in both phases, but the decay rate is singular at the transition 
point: its slope is discontinuous in $d\ge 2$ while in $d=1$ the second 
derivative is discontinuous. 

The averaging over disorder can easily be done exactly and leads to 
\be
   P(c,t) = e^{-\gamma t} (2d)^{-t} Z_t            \label{PZ}
\ee
with 
\be
   Z_t=\sum_{walks}(1-c)^{s}\, b^{x}.                    \label{SAW}
\ee
Here, 
\be
   \gamma=\ln(\frac{b+b^{-1}-2}{2d}+1)
\ee
and $s$ is the number of {\it distinct} sites visited. Finally, $Z_t$ can be 
understood as the partition sum of a {\it self-attracting} polymer (without 
excluded volume interaction; we assume $k_BT=1$) of length $t$, stretched 
by applying opposite forces $\pm \ln b$
to its two ends. The self attraction comes about by the fact that each 
monomer which is placed on a new site is punished by a Boltzmann factor $1-c$,
while no such factor is applied when a monomer is placed at a site which is 
already occupied by another monomer.

Although this is rather artificial as a polymer model, the mapping is very useful 
for suggesting efficient simulation methods and for interpreting the results.

Our simulations are performed using the Pruned-Enriched-Rosenbluth Method 
(PERM) \cite{g97} which is a growth scheme to build weighted walks:
Walks with too small weights are pruned, promising 
configurations are further encouraged by cloning them if their weight exceeds a 
(length-dependent) given value. PERM has been used to simulate efficiently a 
large number of polymer problems including $\Theta$-collapse, DNA denaturation, 
protein folding, and critical unmixing \cite{gn01}. In addition to the 
``population control" provided by cloning and pruning, walks are also 
{\it guided} by assigning different a priori probabilities to different 
jumps. In particular, we chose these such that previously unvisited sites are 
unfavored against previously visited ones, $p_{unvisited}=(1-c)p_{visited}$,
and jumps in the bias direction are more likely than others,
$p_{+x}:p_{-x}:p_{\perp} = \sqrt{b}:\sqrt{1/b}:1$. These a priori biases are 
of course compensated by appropriate weight factors $\propto 1/p$.
 
\section{Unbiased Walkers}

For $b=1$ walks are asymptotically subdiffusive or {\it localized}; the DV 
theory predicts  the mean squared displacement to be $\langle R^{2}\rangle\sim 
t^{2/(d+2)}$. 
However, for small trap concentrations walks are brownian at small 
times ($\langle R^{2}\rangle\sim t$)  reverting to DV scaling asymptotically  
{\it from above} 
\cite{anlauf}. This curious nonmonotonicity (also observed in other 
models of  polymer collapse \cite{gh95,g97,po00} results because the most 
successful 
walks at shorter times do  {\it not} survive for longer periods where DV 
scaling holds (Fig. 1a). The same is seen, albeit much less dramatically, 
in $d=2$ (Fig. 1b). It is not seen in $d=1$ where walks remain localized for 
all times and $\langle R^{2}\rangle$  is monotonic.

\begin{figure}
  \begin{center}
   \psfig{file=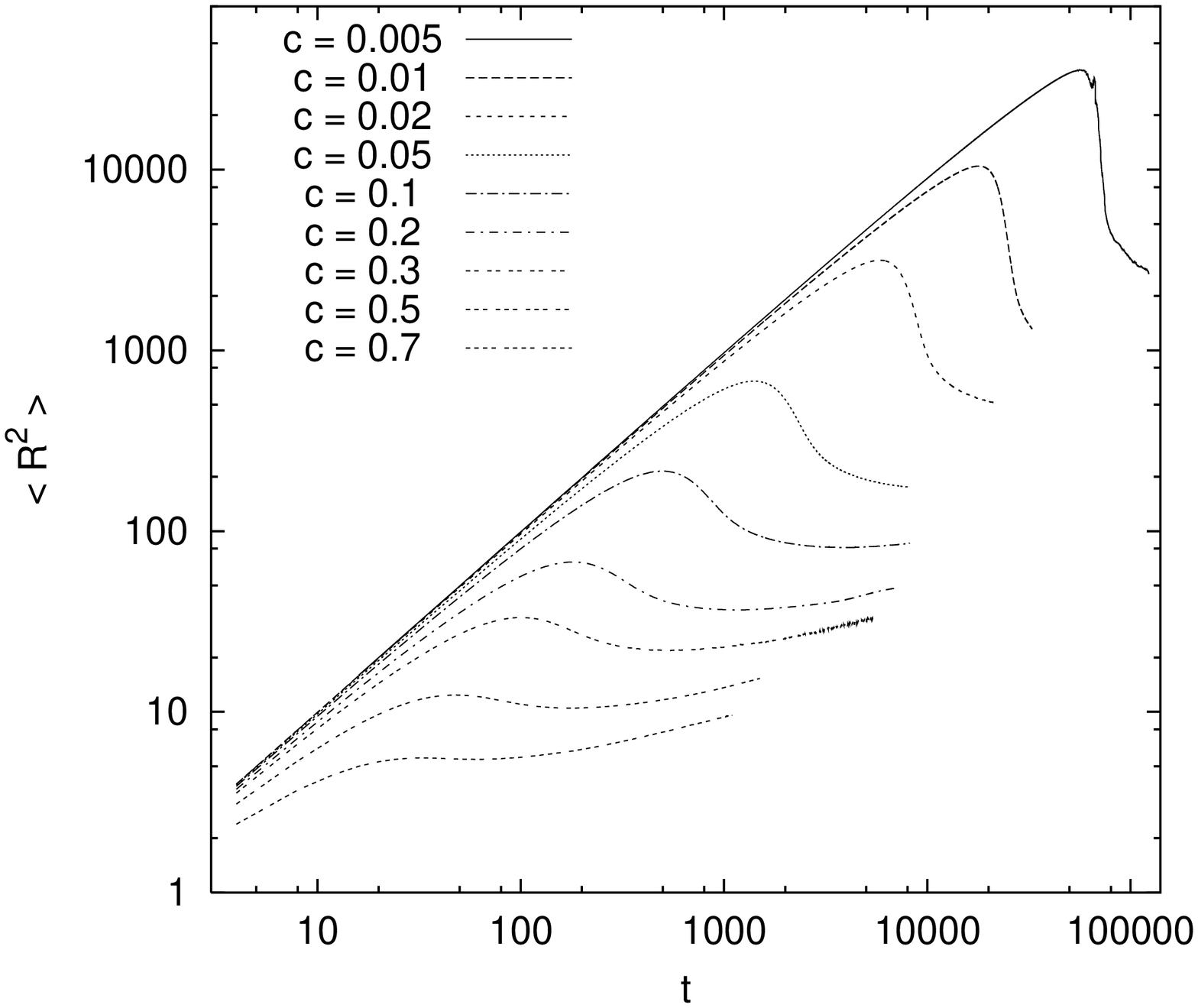,width=3.7cm,angle=270}
   \psfig{file=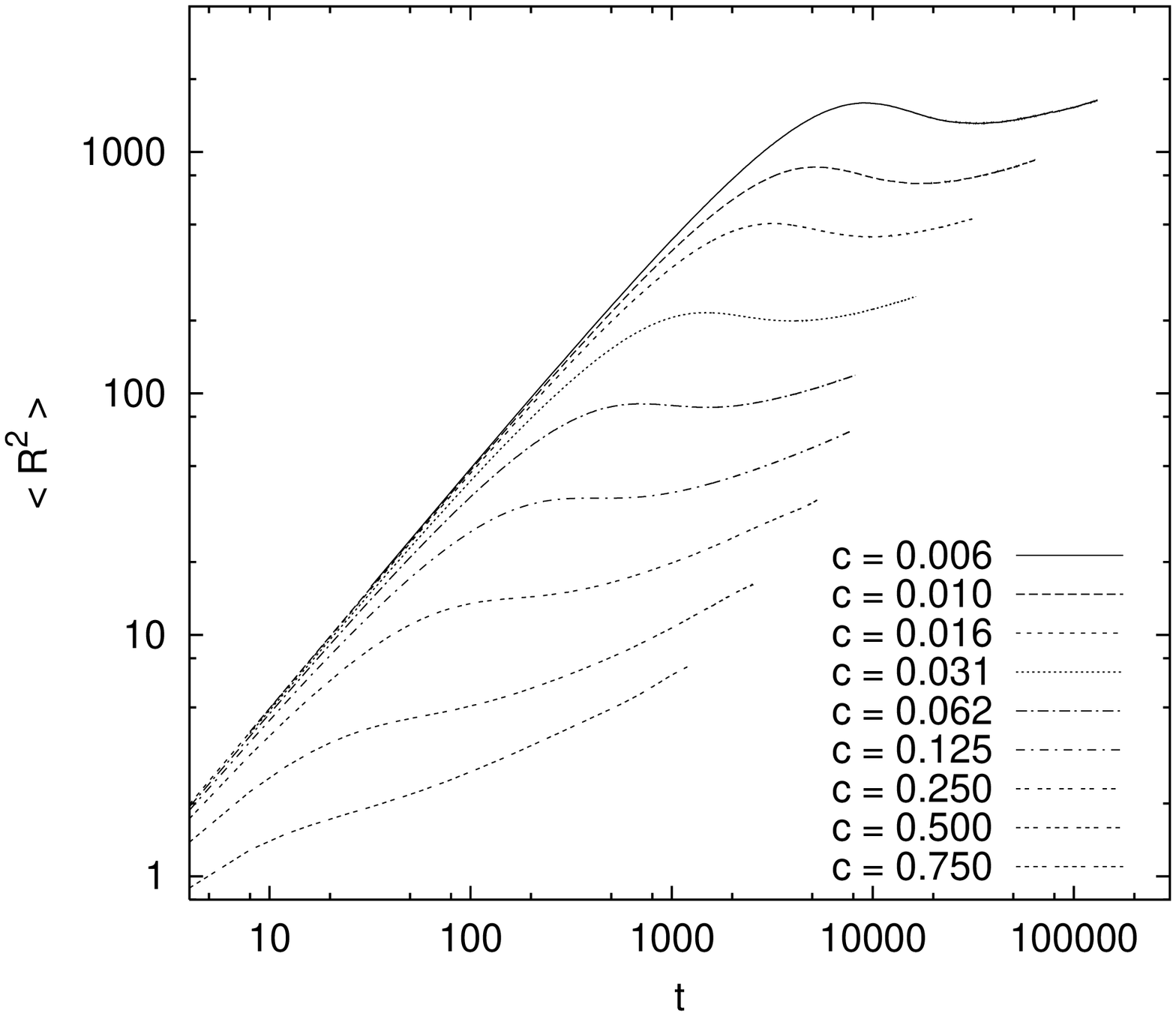,width=3.7cm,angle=270}
   \label{R2_no bias}
   \caption{(a) $\langle R^{2}\rangle$ vs $t$ in $d=3$ for unbiased walks, showing 
      the sudden localization indicated by the drop in $\langle R^{2}\rangle$. The 
      time at which this localization occurs  decreases with 
      increasing trap concentration. The transition proceeds  more gradually for 
      larger $c$, and for $c>=0.7$ no maximum is. (b) Similar data for $d=2$.
      The collapse is there much less pronounced.}
   \end{center}
\end{figure}

The sharp collapse in $d=3$ suggests that the RS-DV crossover is indeed a first
order phase transition in the limits $t\to\infty$ and zero trap 
concentration, as predicted in \cite{n89}.
In the polymer model, it would be a first order transition at infinite 
temperature. A more direct support of this is provided by histograms of the 
distribution of visited sites. For a first order transition we expect two 
peaks for large $t$, the positions of which are separated by a distance $\propto t$.
This is indeed seen in $d=3$ (Fig.2), but not for $d=2$. The latter is not 
surprising in view of the much softer collapse in $d=2$. It could just mean that
the two-peak structure develops only for $t$ values beyond those obtainable by
our simulations ($t\approx 10^5$). 

Combining the RS and DV limits into a single scaling ansatz, Anlauf \cite{anlauf}
suggested for $d=3$ that the average number of visited sites scales as 
\be
   t^{-1} \langle s\rangle = g_3(\lambda t^{2/3}).                         \label{s-scal3}
\ee
A similar ansatz was made for $d=2$ in \cite{gak01}, but as shown in \cite{bbb01} 
the proper ansatz for $d=2$ must involve logarithmic corrections,
\be
   t^{-1} \langle s \rangle \ln t = g_2(\lambda t^{1/2}/\ln t) \;.          \label{s-scal2}
\ee
As seen from Fig.3a, Eq.(\ref{s-scal3}) does not hold with the 
simple scaling function proposed in \cite{anlauf}. Anlauf's data had 
suggested a piecewise linear $g_3(x)$ on a log-log plot, with $g_3(x) = const$
for $x<x^*$ and $g_3(x) = const/x^{2/5}$ for $x>x^*$. We see instead from 
Fig.3a that $g_3(x)$ develops a kink with infinite slope at $x=x^*$. This 
is not too surprising. While a behavior as claimed in \cite{anlauf} would 
be natural for a second order transition, a more singular behavior is expected 
for a first order transition. For $d=2$ the scaling function is much smoother
(see Fig.3b), 
\linebreak[4]

\vglue -7pt
\begin{figure}
  \begin{center}
   \psfig{file=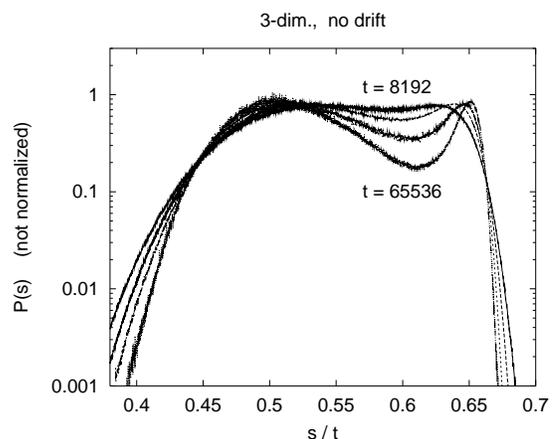,width=6.cm,angle=270}
   \label{P(s)_no bias}
   \caption{Histogram of visited sites for unbiased walks in $d=3$. 
	The two peaks correspond to the collapsed (DV, left) and extended
        (RS, right) phases. Trap concentrations (resp. monomer interaction
        strenghts) have been choosen such that both peaks have equal heights:
     $c = .01993\;(t=8192),\;.01283$
      $(16384),\;.00823\;(32768),\;.00527\;(65536)$.}
   \end{center}
\end{figure}

\noindent
although this might be a transient effect only. We finally 
point out that we require the precise asymptotics for 2-d random walks as derived 
in \cite{hs92}, with $\ln t$ replaced by $\ln (8t)$, to obtain a good data 
collapse.

\begin{figure}
  \begin{center}
   \psfig{file=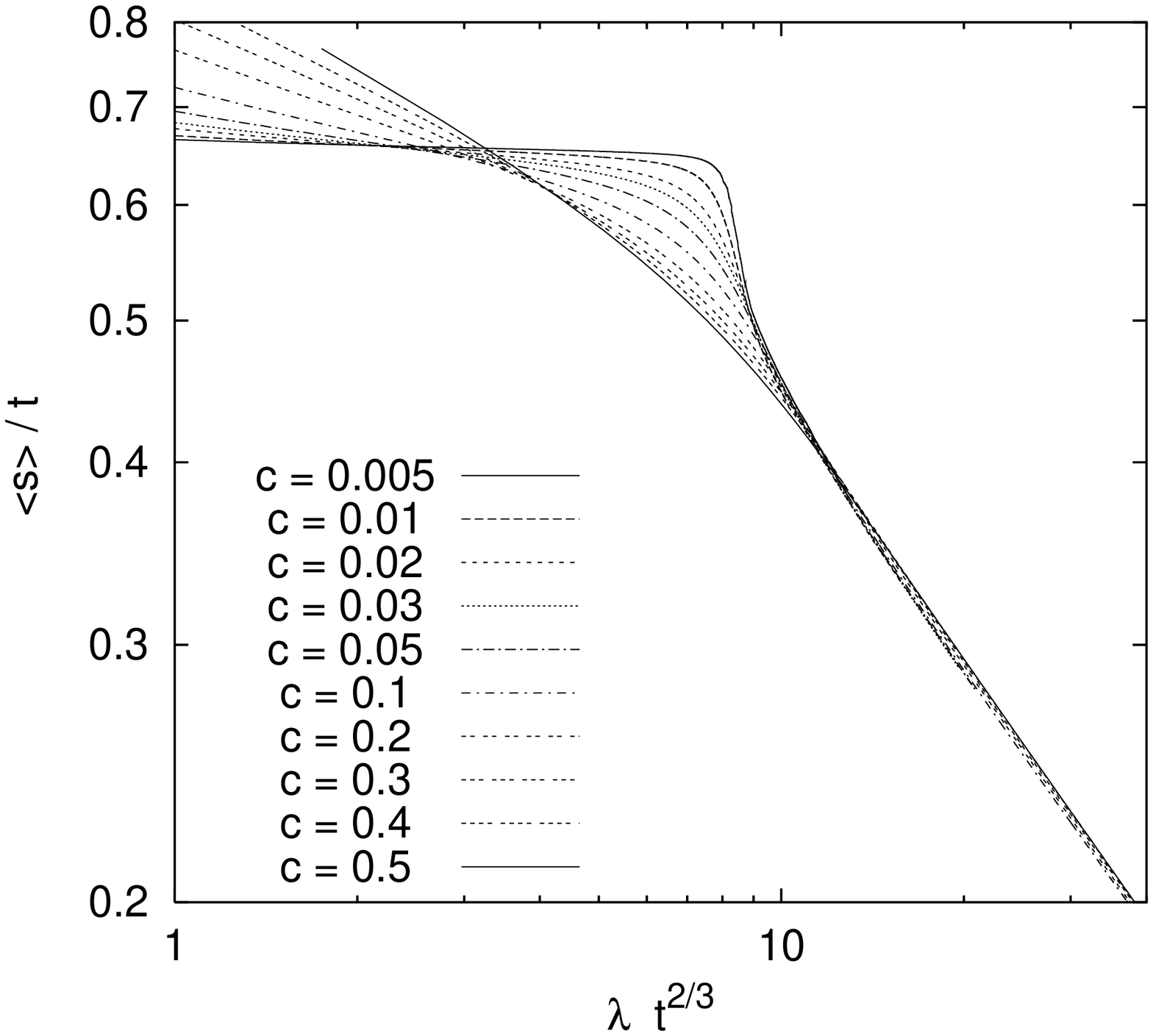,width=3.7cm,angle=270}
   \psfig{file=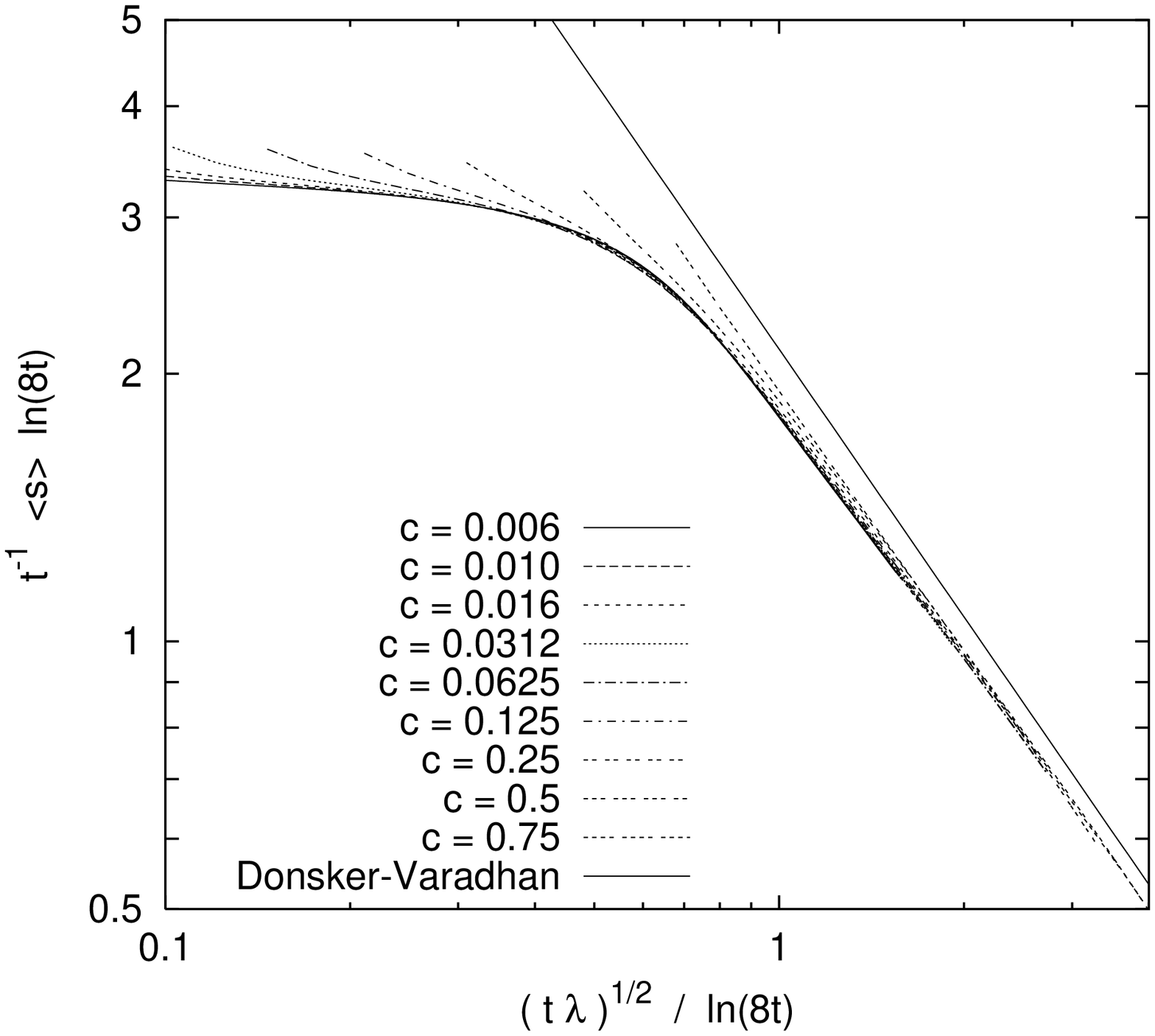,width=3.7cm,angle=270}
   \label{av_s_nobias}
   \caption{(a) $\langle s\rangle / t$ vs $\lambda t^{2/3}$ in $d=3$ for unbiased 
      walks. For large $t$ the data collapse onto a scaling function which is 
      flat for small arguments, $\propto (\lambda t^{2/3})^{-2/5}$ for large 
      arguments, and has a cusp with infinite right slope in between. It is 
      not clear from the data whether the scaling function is continuous or 
      discontinuous at the cusp. (b) Similar for $d=2$. Notice that we replaced 
      $\ln t$ in Eq.(6) by $\ln (8t)$ in order to improve the data collapse. The
      straight line is the asymptotic DV prediction.}
   \end{center}
\end{figure}

\vglue -12pt
\begin{figure}
  \begin{center}
   \psfig{file=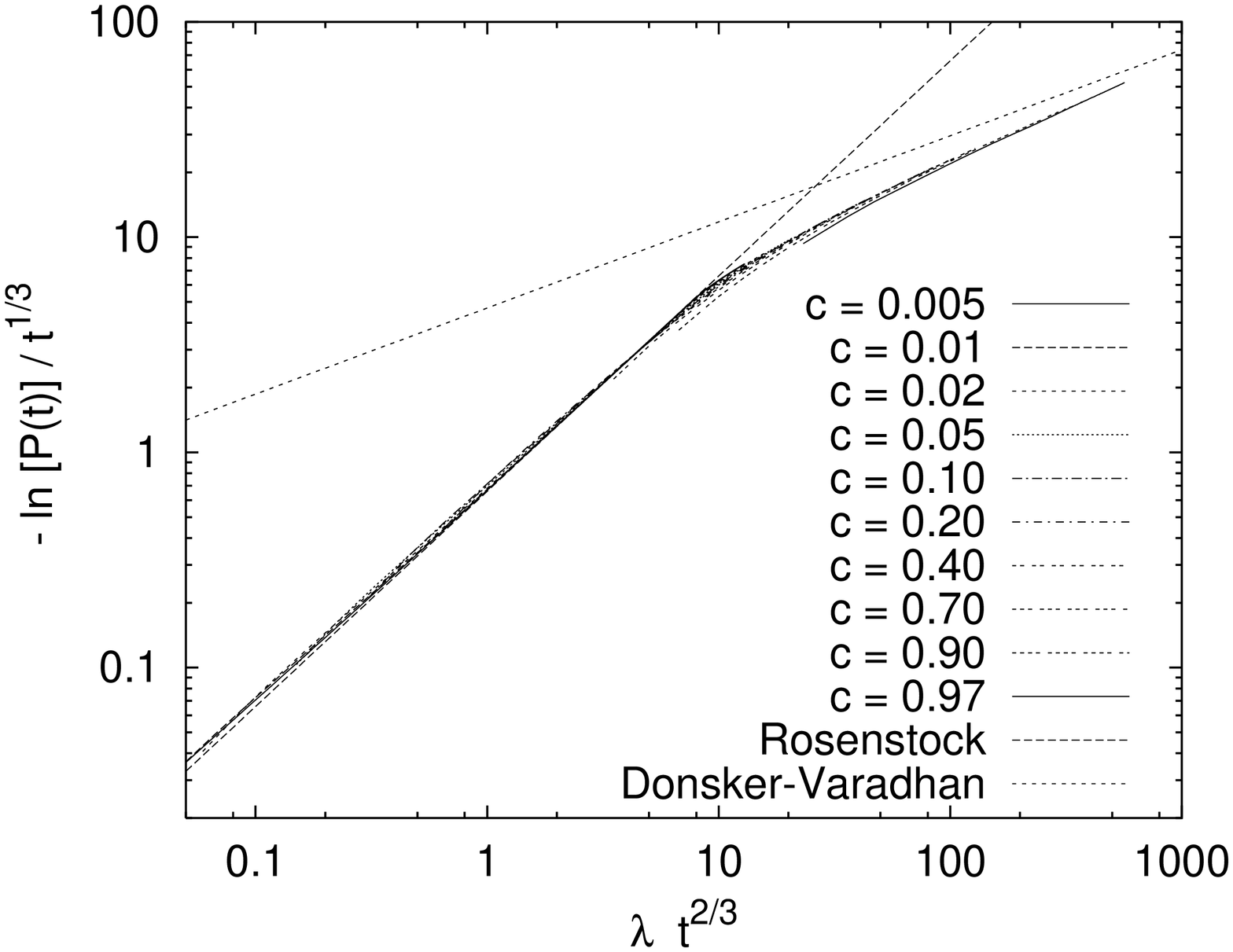,width=4.7cm,angle=270}
   \psfig{file=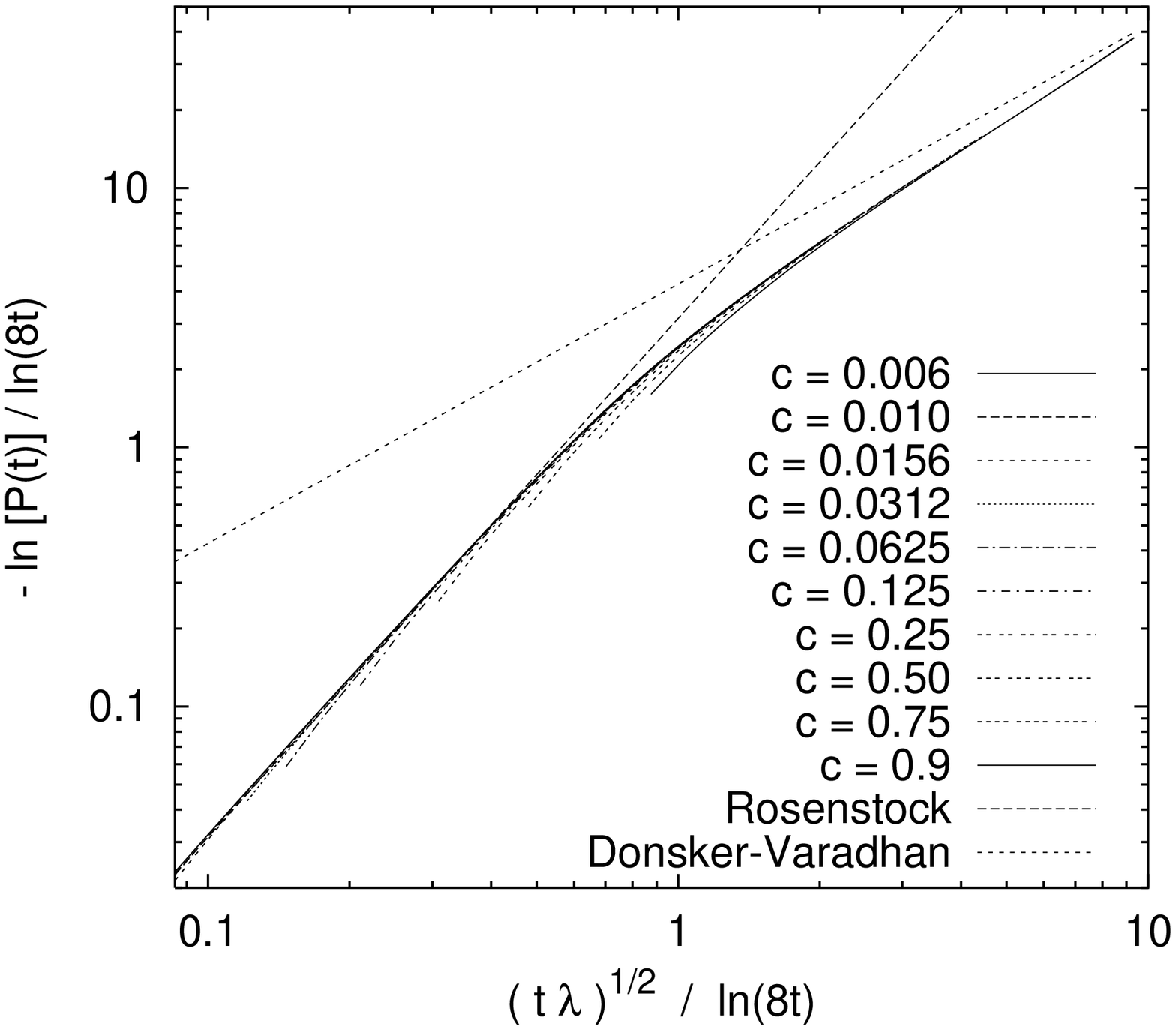,width=4.7cm,angle=270}
   \vglue 10pt
   \label{P_nobias}
   \caption{(a) Scaling plot for $P(c,t)$ for unbiased walks in $d=3$. The straight
      lines are the RS and DV predictions. While the RS prediction is realized 
      numerically (for small concentrations $c$) up to the transition point, the 
      true DV scaling is reached only very slowly. (b) Similar for $d=2$.}
   \end{center}
\end{figure}

Using Eq.(\ref{s-scal3}), Anlauf had proposed a similar scaling for $P(c,t)$,
\be
   t^{-1/3}\ln P(c,t) =f_3(\lambda t^{2/3}).                       \label{scal3}
\ee
Using his simple $g_3(x)$ he derived, by thermodynamic integration, an $f_3(x)$
which gave a very slow convergence to the asymptotic DV scaling after having deviated 
sharply and suddenly 
from RS behavior. Although we disagree with his 
scaling function, our results for $P(c,t)$ (Fig.4a) fully confirm with his 
predictions within the numerical errors. Notice that this disagrees slightly 
with the results found in \cite{bbb01}. Again the analogous ansatz for $d=2$ 
should involve logarithms \cite{bbb01}, and for a faster convergence we must 
replace $\ln t$ by $\ln (8t)$,
\be
   \ln P(c,t) /\ln (8t) = f_2(\lambda t^{1/2}/\ln (8t)) \;.        \label{scal2}
\ee
Agreement with simulations is very good, and again we see very slow 
convergence to the DV asymptotics, in contrast to \cite{bbb01}.

\section{Nonzero Bias}

For any nonzero bias the long time probability decay is exponential \cite{gp82a,s87}. 
This is most easily seen as follows: Let $\phi(x,{\bf x}_\perp,t;b;\cal{C})$ 
be the probability density, for any fixed trap configuration $\cal{C}$ and bias $b$,
that a walker who started at $t=0$ in the origin ${\bf x}=0$ is at time $t$ at the 
position ${\bf x} = (x,{\bf x}_\perp)$. One verifies easily that this is related 
to the analogous distribution of the bias-free case by (see \cite{a86} for the 
continuum case)
\be
   \phi(x,{\bf x}_\perp,t;b;{\cal C}) = e^{-\gamma t} b^x 
              \phi(x,{\bf x}_\perp,t;1;{\cal C}) \;.               \label{gauge}
\ee
This `gauge transformation' means in particular that the return probability
\be
   P_{\rm return}(t) = \langle \phi(0,{\bf 0},t;b) \rangle_{\cal C}
\ee
decays exactly as 
\be
   P_{\rm return}(t) = e^{-\gamma t} P_{\rm return}^{(b=1)}(t)\;,
\ee
i.e. as an exponential multiplied by the DV stretched exponential since $P_{\rm return}
^{(b=1)}(t)$ also satisfies the DV estimate up to algebraic factors. 

For the survival probability we obtain
\be
   \ln P(c,t) = \ln P^{(b=1)}(c,t) - \gamma t + 
          \ln \langle \cosh(x\ln b) \rangle_{{\rm walks},\,\cal C}\;.
                                                                 \label{decay}
\ee
Since the last term is positive definite, we have the exact inequality
$\ln P(c,t) \ge \ln P^{(b=1)}(c,t) - \gamma t$. Moreover, if the walkers 
stay localized in spite of the bias, $x$ increases sublinearly with $t$ and 
thus the asymptotic decay rate is
\be
   \alpha \equiv - \lim_{t\to\infty} {1\over t} \ln P(c,t) = 
          \gamma\;. \qquad {\rm (localized)}                     \label{local-decay}
\ee
Notice that this is independent of the trap concentration.
A smaller decay rate can be obtained only when the walkers drift, but then 
the chance to hit a trap is finite at any time step, and $P(c,t)$ decays 
again exponentially. Indeed, in the drifting regime any correlations should be 
short, and thus both the decay rate and the drift velocity are attained 
already after a short time.

\subsection{$d=1$}

In one dimension it is possible to derive  analytic estimates for the survival 
probability and the drift velocity of delocalized walks. An exact lower bound 
on the survival probability in terms of probabilities of large trap-free regions 
is obtained in a way similar to \cite{gp82}, by estimating the contributions 
of trap-free regions of length $l$ downstream of the starting point of the 
walker. Take some positive integer $l$ and consider configurations with no trap 
in the interval $[0,l]$. We then have 
\be
   P(c,t) \ge (1-c)^l \; P_l(t)                                 \label{P-1d}
\ee
where the first factor on the rhs. is the probability to find such a gap in the 
trap configuration, and $P_l(t)$ is the probability to survive in it. The latter 
can be estimated using Eq.(\ref{gauge}), the method of images to incorporate 
the boundary conditions at $x=-1$ and $x=l+1$, and Stirling's formula. We 
obtain, up to a constant dependent only on $b$, 
\be
   \ln P_l(t) \ge l\ln b - \gamma t -
           \left[ {t+l\over 2}\ln(1+{l\over t})+{t-l\over 2}\ln(1-{l\over t})\right].
\ee
The rhs. of Eq.(\ref{P-1d}) is maximal for $l=l^*\equiv t \tanh[\ln((1-c)b)]$
when $b > b_c \equiv 1/(1-c)$, and for $l = o(t)$ when $b<b_c$. Thus the 
decay rate is bounded by $\alpha \le \gamma - \ln[\cosh(\ln(1-c)b))]$ for $b>b_c$. 
Actually we claim that this bound is saturated, since it can also be understood 
as the result of a (non-rigorous) saddle point approximation. Summarizing, we 
have thus found that the decay rate is equal to $\gamma$ for $b<b_c$ and is
given by $\alpha = \gamma - \ln[\cosh(\ln(1-c)b))]$ for $b>b_c$. Similarly, the 
drift velocity of the survivors, defined as $v=\lim_{t\to\infty}
\langle x\rangle /t$, vanishes for $b<b_c$ and is given 
by $v = \tanh[\ln((1-c)b)]$ for $b>b_c$. In the continuum limit
$(c\to 0,\; (1-c)b = finite)$ this agrees with \cite{rk84,a86,movaghar}, but for
finite $c$ it disagrees with \cite{rk84}\cite{footnote2}. Thus there
is a phase transition at $b=b_c$. It is second order
in the sense that $v, \alpha$, and $d\alpha/db$ are continuous at $b=b_c$, while
$dv/db$ and $d^2\alpha/db^2$ are discontinuous. 

These predictions are fully
confirmed by simulations, as seen from Figs.5 and 6. In Fig.5 we see $\ln P(c,t) +
\gamma t$ versus $t$ for $c=0.1$ and for several values of $b$. First of all we 
see that all curves are above that for $b=1$, as predicted theoretically. 
Secondly, we see that all curves 
for $b\le 1.11$ become horizontal for $t\to
\infty$, in agreement with the prediction $b_c = 1/(1-c) = 1.111\ldots$. The 
decay rates obtained from Fig.5 for $b>b_c$ are shown in Fig.6a, together with 
the theoretical predictions. We see that the transition is indeed smooth, as 
obtained theoretically. Finally, the drift velocities for $c=0.1$ are shown 
\linebreak[4]

\vglue -10pt
\begin{figure}
  \begin{center}
   \psfig{file=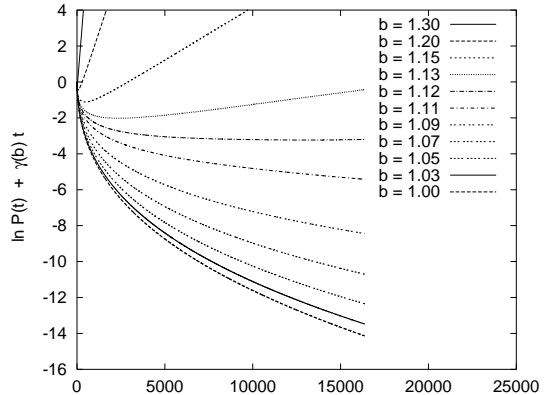,width=5.5cm,angle=270}
   \label{P-drift-1d}
   \caption{$\ln P(c,t) + \gamma t$ for $d=1$ and trap concentration $c=0.1$.
     Statistical errors are smaller than the thickness of the curves.}
   \end{center}
\end{figure}

\vglue -14pt
\begin{figure}
  \begin{center}
   \psfig{file=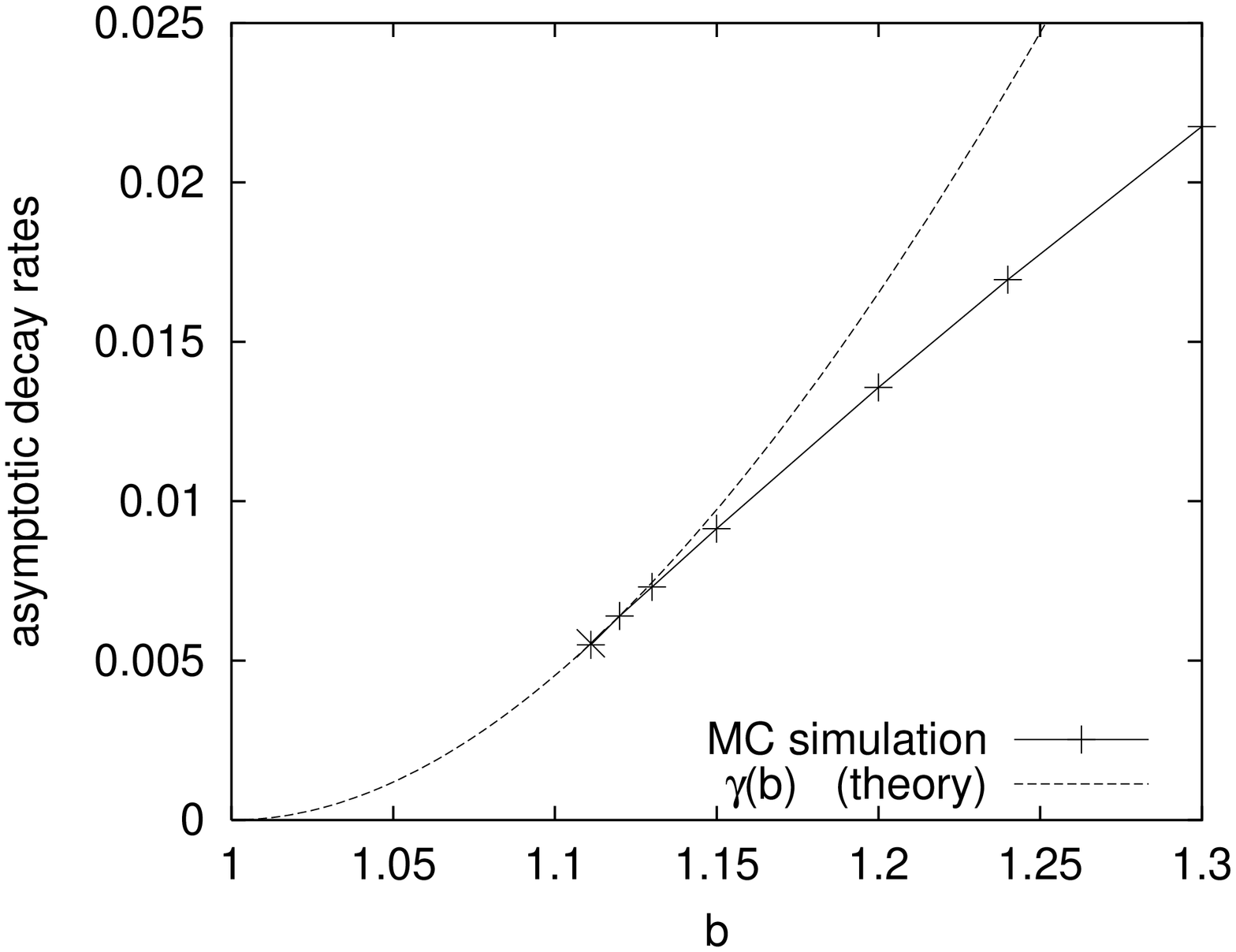,width=4.7cm,angle=270}
   \psfig{file=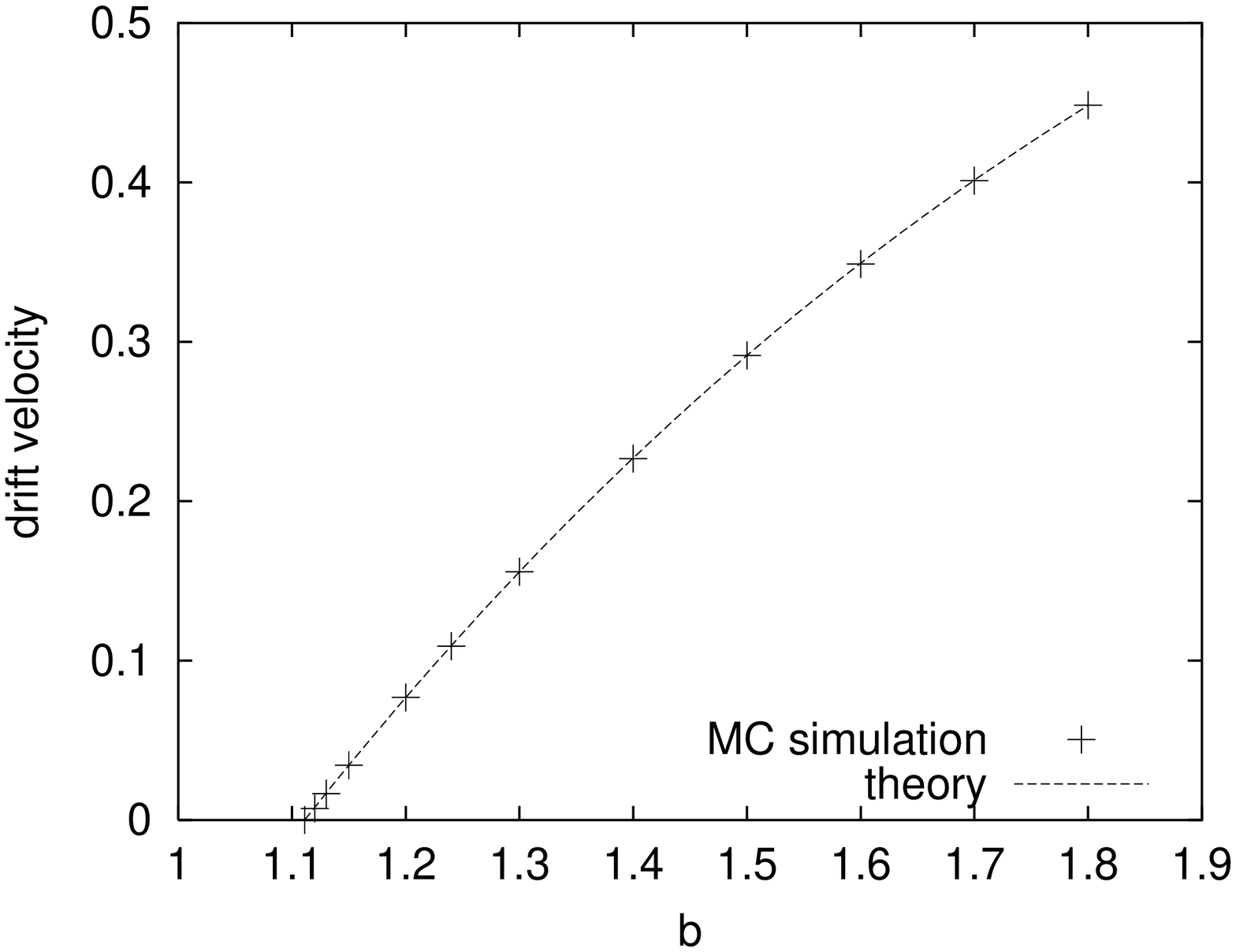,width=4.7cm,angle=270}
\vglue 5pt
   \label{drift-1d}
   \caption{(a) Decay rate $\alpha$ as a function of bias $b$, for $c=0.1$ ($d=1)$. For 
     $b\ge 1.2$ the statistical errors are much smaller than the size of the symbols,
     for smaller $b$ there is an increasing systematic uncertainty due to the 
     non-trivial extrapolation for $t\to\infty$. 
     (b) Analogous results for the drift velocity.
     Dashed curves are analytic predictions.}
   \end{center}
\end{figure}

\noindent
in Fig.6a. Again agreement with theory is perfect. We should add that similar 
results were also found for other values of $c$, and that $\langle x\rangle$
increased always monotonically with $t$, in contrast to what is found for 
$d>1$ (see below).

\subsection{$d\ge 2$}

In higher dimensions we don't have similar analytic results. Moreover, direct 
simulations of the transition region are very difficult, even with our efficient
algorithm. But Eq.(\ref{local-decay}) together with the expected fast convergence 
of the decay rate and drift velocity in the delocalized phase allow us, in spite
of this difficulty, to compute $v, \alpha$, and $b_c$ with very high precision.

In Fig.7 we show the average displacements $\langle x\rangle$ in $d=2$ against $t$,
for $c=1/8$ and for several values of $b$. Very similar data are obtained for $d=3$ 
and for other values of $c$. 
Obviously $b_c > 1.65$, since for all $b \le 1.65$ the curves decrease after an initial 
rise $\langle x\rangle \propto t$. The reason for this non-monotonic behavior 
is the same as in Fig.1: Walkers who venture far out do better initially, but 
finally only those win who started in a large trap-free region and stayed in it.
From Fig.7 one might guess that $1.65 < b_c < 1.7$, but this would be wrong: 
Also for 
$b=1.7$ the curve should bend down ultimately, but this will happen very late 
(see below) 
and our algorithm might easily miss it (in the transition region it generates
many drifting walks with low weight and only very few compact walks, but these with
high weight).

The true transition point can be estimated from the survival probabilities, 
shown in Fig.8 for $d=2$. Actually, in view of eq.(\ref{decay}), we plot 
$\ln P(c,t) + \gamma t$. For small $b$ we see curves which follow closely 
the curve for $b=1$, but are slightly above it. These curves contain a linear 
part at small $t$ whose slope decreases and whose length increases with $b$.
The cross-over from this linear part to the part following the DV curve is 
rather sharp and stays so with increasing $b$. Since the DV curve is asymptotically 
flat, this scenario should hold up to the value of $b$ where the ``initial" part becomes 
horizontal. Beyond that value we are in the drifting phase. Obviously
the initial straight parts in Fig.8 correspond precisely to the (transient)
drifting phase seen already in Fig.7. Indeed, the cross-overs between the 
two regimes are at the same positions in Figs. 7 and 8.

This means that $b_c$ is implicitly given by the condition $\alpha_{\rm drifting} = \gamma$,
where $\alpha_{\rm drifting}$ is the decay rate in the drifting phase. 
While this phase is asymptotically stable for $\alpha_{\rm drifting} < \gamma$,
it is only transient for $\alpha_{\rm drifting} > \gamma$. Even in the 
transient regime it can be measured precisely (together with the drift velocity), 
if this regime lasts long enough. Moreover, the drifting phase exists also 
for $b<b_c$ and large $t$, but is only metastable in this regime. This is 
clearly seen from 
histograms of the $x$-distribution. In Fig.9 (which refers 
to $d=2$ and $c=1/8$) we see clearly two peaks in $P(x)$, for $t\ge 900$. The 
left peak is at $x/t\approx 0$. Indeed, it is slightly to the right of $x/t=0$, 
but it moves to $x/t=0$ for $t\to\infty$. It of course corresponds to the 
localized phase. The right peak is the drifting phase. The fact 
\linebreak[4] 

\begin{figure}
  \begin{center}
   \psfig{file=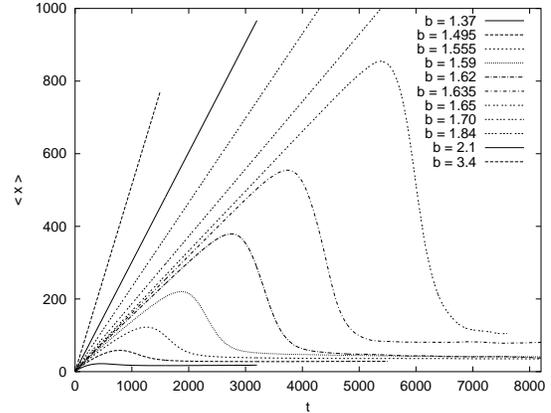,width=5.4cm,angle=270}
   \label{2d-xav_drift}
   \caption{Average displacement in the bias direction for $c=1/8$ in $d=2$.}
   \end{center}
\end{figure}

\begin{figure}
  \begin{center}
   \psfig{file=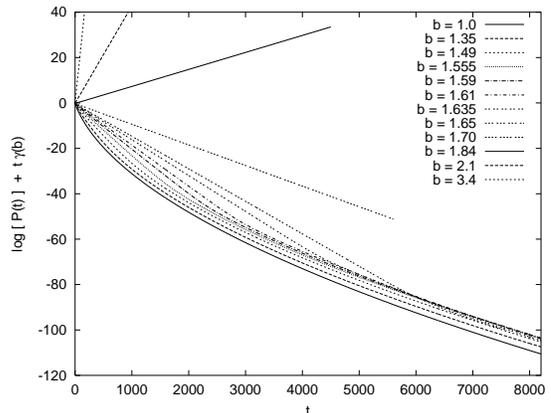,width=5.4cm,angle=270}
   \label{2d-P_drift}
   \caption{$\ln P(c,t) + \gamma t$ for $d=2$ and trap concentration $c=1/8$. Since 
     these curves must be increasing with $b$, the fact that the curve for $b=1.65$
     crosses for $t>6000$ below those for $b=1.555$ to 1.635 is due to numerical 
     inaccuracy.}
   \end{center}
\end{figure}

\noindent
that these 
peaks become sharper and move apart with increasing $t$ shows that the 
transition is {\it first order}. For $d>2$ we expect the transition then to 
be of first order {\it a fortiori}.

Since $\gamma$ is known analytically, 
we can locate $b_c$ by simulations in the delocalized phase alone. 
This procedure is indicated in Fig.10 where the dashed curve is $\gamma(b)$
and the various continuous curves are {\it finite-time} decay rates obtained from
{\it short-time} runs with different trap concentrations. The error bars of 
the latter are much smaller than the size of the symbols. The intersection 
points give $b_c$ and $\alpha_c$ in terms of the trap concentration $c$. 

The scaling behavior for $c\to 0$ is easily obtained by observing that 
$\gamma \approx (b-1)^2/(2d)$ for $b\approx 1$, while $\alpha \sim c$ in
the drifting phase (see Fig.11). 
The latter is true for small $c$ since the Rosenstock approximation 
becomes exact in this limit. Therefore we predict that $b_c-1 \sim \sqrt{c}$. 
This is verified in Fig.12 for $d=2$ and $d=3$. Finally, from Fig.13 we 
see that also $v_c \sim \sqrt{c}$, i.e. $v_c \propto b_c$. The latter is 
very natural, but we must remember that it is not true in $d=1$.

\begin{figure}
  \begin{center}
   \psfig{file=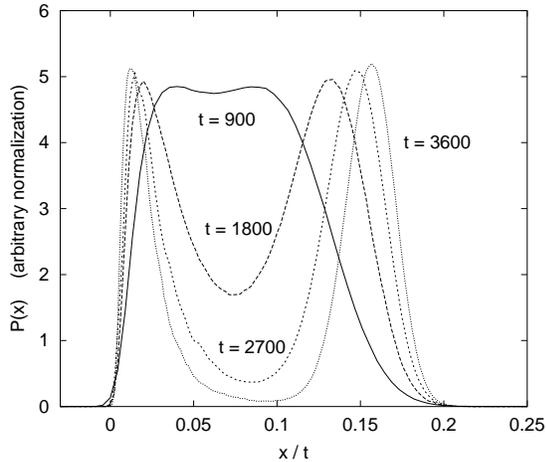,width=6.3cm,angle=270}
   \label{2d-histo_drift}
   \caption{Histograms of the rescaled end point distance $P(x/t)$ versus $x/t$
     for $d=2$ and $c=0.125$. Bia\-ses were adjusted so that both peaks have 
     equal height: $b=1.510\; (t=900),\; 1.577\; (t=1800),\; 1.608\; (t=2700),\; 1.628$
     $(t=3600)$. Normalization is arbitrary. The peak at $x/t \approx 0$ is due to
     the localized phase, the other corresponds to the drifting phase.}
   \end{center}
\end{figure}

\begin{figure}
  \begin{center}
   \psfig{file=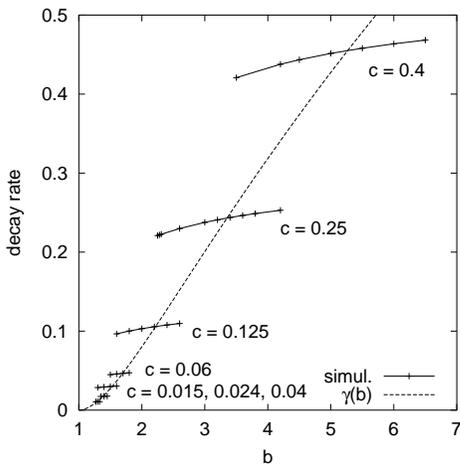,width=6.3cm,angle=270}
   \label{3d-P_drift}
   \caption{Decay rates $\alpha$ measured in the drifting (delocalized) phase, $d=3$.
     The dashed curve is the prediction $\alpha=\gamma(b)$ for the localized 
     phase. To the left of this curve, the drifting phase is only transient. 
     In spite of that, $\alpha$ is easily measured with high precision, thus the 
     statistical errors are $\le 10^{-4}$, much smaller than the symbols.}
   \end{center}
\end{figure}

\begin{figure}
  \begin{center}
   \psfig{file=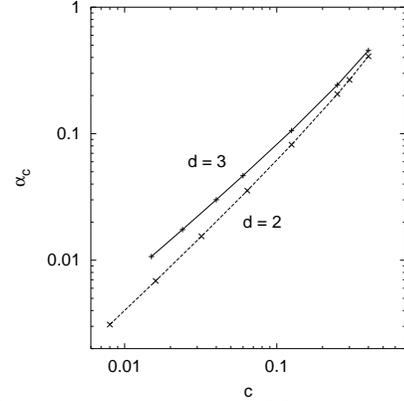,width=5.5cm,angle=270}
   \label{alpha-versus-c}
   \caption{Log-log plot of the critical decay rate versus $c$.
     For small $c$, both curves scale $\propto c$.}
   \end{center}
\end{figure}

\begin{figure}
  \begin{center}
   \psfig{file=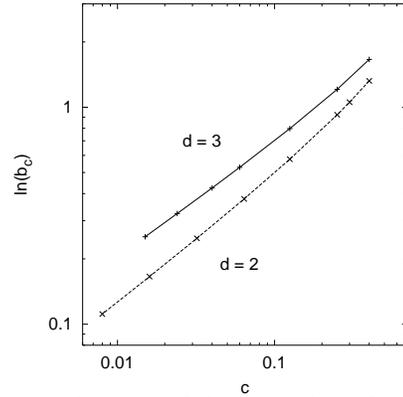,width=5.5cm,angle=270}
   \label{3d--bias_v-concentr}
   \caption{Log-log plot of the logarithm of the critical bias versus $c$. 
     For small $c$, both curves scale as $\sqrt{c}$.}
   \end{center}
\end{figure}

\begin{figure}
  \begin{center}
   \psfig{file=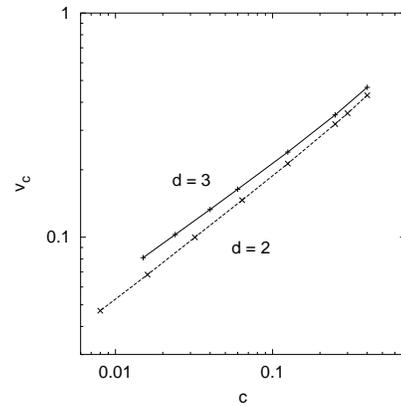,width=5.5cm,angle=270}
   \label{v-versus-c}
   \caption{Log-log plot of the critical drift velocity versus $c$.}
   \end{center}
\end{figure}

\section{Alternative Interpretations}

\subsection{Population Dynamics}

Up to now, we dealt with independent walkers. Detailed studies of this model 
were often criticized on the grounds that the tiny probabilites that obtain
in the DV regime render the results physically irrelevant 
(in the model with bias, probabilities are 
even smaller). This criticism can be countered by adding autocatalytic 
particle production (local production of off-springs) \cite{nelson}. 

Thus we consider a 
model where particles can hop, get absorbed when hitting a trap, and can 
reproduce with some fixed rate $\sigma$. For simplicity we do not include
any interaction between particles, in particular we do not take into account 
any excluded volume effect or any competition for food or other resources.
The only effect of reproduction is then that the average number of survivors 
at generation $t$ is $P(c,t) (1+\sigma)^t$. 

In the bias-free case any non-zero $\sigma$ will thus lead to an explosion 
of the population, indicating that the model is basically sick. But in 
the case with drift a suitably chosen rate $\sigma$ will lead to either 
a much reduced decay of the particle number, or even to a stationary 
population size. In this case the neglect of excluded volume and saturation
effects will be much more benign, and the effects studied in this paper 
might have a bigger chance to be observable in real experiments.

Finally, including excluded volume effects would lead to directed 
percolation (contact process) in a random medium \cite{noest}.

\subsection{Stretched Collapsed Polymers}

We have already pointed out that $Z_t$ (Eq.(\ref{SAW})) can be understood as 
the partition sum of a polymer with self-attraction of strength $k_BT\,\lambda$,
and stretched by a force $k_BT\,\ln b$. The decay rate is then essentially 
the free energy per monomer in the thermodynamic limit. 

The fact that unfolding of a collapsed polymer is a first order transition 
is well known \cite{zhulina,wittkop,lai}. The coexistence 
of stretched and coiled phases corresponds to a ``tadpole" configuration 
\cite{ball,zhulina} where part of the chain is collapsed and the rest sticks 
out of the coil, so that the total length $x$ is equal to some prescribed 
value. Our finding that the critical bias is $\propto \sqrt{c}$ means then 
that the string tension in a tadpole configuration scales, in the 
limit of a very long chain and close to the collapse point, as the square root 
of the monomer-monomer attraction strength. This seems to be a new result,
and it remains to see whether it holds also for more realistic polymer models.

The main difference to standard polymer models is the fact that we 
have neglected here excluded volume effects (they have no counterpart in the 
trap problem). Thus the collapse of the unstretched polymer happens at 
infinite temperature in the present model. In the standard polymer models 
with next-nearest neighbor attraction or with three body repulsion the 
$\Theta$-collapse is second order, in contrast to the present model. But the
dependence of $\langle R^2\rangle$ on chain length is also non-monotonic in
$d\ge 3$ \cite{gh95,g97,po00}. In $d=4$ the collapse is indeed ``quasi first 
order" \cite{po00}, and special care is needed to realize that it actually is 
continuous. We believe that in the present model the transition is truely of
first order.

\subsection{Magnetic Flux Lines}

A last interpretation is in terms of magnetic flux lines in a type II 
superconductor with parallel columnar defects \cite{nelson}. These defects are
quenched and randomly distributed. They tend to pin the flux lines,
thus preserving superconductivity in the sample. In the unbiased 
case, the mean orientation of the flux lines is parallel to the defects, 
caused by a magnetic field in the same direction. 
The biased case corresponds then to a non-zero angle between the field and 
the defects. For small magnetic field the flux lines will essentially follow 
the defects. But above a critical field strength (resp. tilting angle) the 
average angle between flux lines and defects is non-zero. Our main result is 
that this transition is first order in the sense that the average angle jumps 
by a finite amount when passing through the critical field strength. This jump
scales as the square root of the defect density.

\section{Conclusion}

Using extensive Monte Carlo simulations by means of the PERM algorithm, we 
have studied the problem of particles performing unbiased and biased 
random walks in media with randomly located traps. In the bias free case
we find the first clear numeric indication for the first order nature of 
the Rosenstock - Donsker-Varadhan transition in 3 dimensions. In $d=2$ 
the situation is less clear, but we showed that also there the transition 
is rather sharp. When the diffusion is biased, we find a transition from 
localized to drifting walkers at a critical bias which depends on the trap
density. This transition is second order in $d=1$ but first order in $d\ge 2$.
Combining the simulations with exact analytic estimates we were able to 
draw the phase diagram with high accuracy.

Acknowledgements:

We are indebted to Walter Nadler for numerous discussions and for critically reading 
the manuscript, and to Gerard Barkema for correspondence.

\end{multicols}


\begin{thebibliography}{99}
\bibitem{footnote} Actually, this ``velocity" is the defined as the change $d
   \langle x\rangle / dt$, where $\langle x\rangle$ is the average position of 
   the survivors. Since walkers get absorbed, this is not equal to the average 
   velocity of individual walkers.
\bibitem{footnote2} This seems to be due to an unjustified Gaussian approximation
   to Eq.(15) made in \cite{rk84}, $\ln P_l(t) \approx l\ln b - \gamma t -l^2/(2t)$.
\end{thebibliography}
\end{document}